\documentclass[journal]{IEEEtran}
\usepackage{stfloats}
\usepackage{float}
\usepackage{graphicx}
\usepackage{subfigure}
\usepackage{color}
\usepackage{algpseudocode}
\usepackage{amssymb}
\usepackage{amsmath}
\usepackage{algorithm}
\usepackage{bm}
\usepackage[colorlinks]{hyperref}

\begin{document}

\title{\LARGE FAS-RIS Communication: Model, Analysis, and Optimization}

\author{Junteng Yao, Jianchao Zheng, Tuo Wu, Ming Jin,  Chau Yuen, \emph{Fellow, IEEE},\\
Kai-Kit Wong, \emph{Fellow}, \emph{IEEE},  and Fumiyuki Adachi, \emph{Life Fellow}, \emph{IEEE}
\vspace{-9mm}

\thanks{(\textit{Corresponding author: Tuo Wu.})}
\thanks{J. Yao and M. Jin are with the Faculty of Electrical Engineering and Computer Science, Ningbo University, Ningbo 315211, China (E-mail:  $\rm \{yaojunteng, jinming\}@nbu.edu.cn$).}
\thanks{J. Zheng is with the School of Computer Science and Engineering, Huizhou University, Huizhou 516000, China (E-mail: $\rm zhengjch@hzu.edu.cn$).}
\thanks{T. Wu and C. Yuen are with the School of Electrical and Electronic Engineering, Nanyang Technological University, 639798, Singapore (E-mail: $\rm \{tuo.wu, chau.yuen\}@ntu.edu.sg$).}
\thanks{K.-K. Wong is with the Department of Electronic and Electrical Engineering, University College London, WC1E 6BT London, U.K., and also with the Yonsei Frontier Laboratory and the School of Integrated Technology, Yonsei University, Seoul 03722, South Korea (E-mail: $\rm kai\text{-}kit.wong@ucl.ac.uk$).}
\thanks{F. Adachi is with the International Research Institute of Disaster Science (IRIDeS), Tohoku University, Sendai, Japan (E-mail: $\rm adachi@ecei.tohoku.ac.jp$).}
}

\maketitle

\begin{abstract}
This correspondence investigates the novel fluid antenna system (FAS) technology, combining with reconfigurable intelligent surface (RIS) for wireless communications, where a base station (BS) communicates with a FAS-enabled user with the assistance of a RIS. To analyze this technology, we derive the outage probability based on the block-diagonal matrix approximation (BDMA) model. With this, we obtain the upper bound, lower bound, and asymptotic approximation of the outage probability to gain more insights. Moreover, we design the phase shift matrix of the RIS in order to minimize the system outage probability. Simulation results confirm the accuracy of our approximations and that the proposed schemes outperform benchmarks significantly.
\end{abstract}

\begin{IEEEkeywords}
Fluid antenna system (FAS), outage probability, reconfigurable intelligent surface (RIS).
\end{IEEEkeywords}

\vspace{-2mm}
\section{Introduction}
\IEEEPARstart{W}{ith the rise} of Internet-of-Things (IoT), massive connectivity of sensors and small communication devices is becoming a pressing demand but their affordability to space and complexity greatly obstructs network performance. To overcome this bottleneck, fluid antenna system (FAS) has emerged as a promising solution \cite{Wong-fas-cl2020}. FAS is motivated by new forms of reconfigurable antenna to enable shape-flexible and position-flexible antenna technologies \cite{wong2023fluid}. Recent experiments on FAS have been reported in \cite{Shen-tap_submit2024,Zhang-pFAS2024}.

In FAS, the radiating position (a.k.a.~port) can be optimized over a prescribed space, providing a new degree-of-freedom (DoF) for optimization in the physical layer \cite{KKWong21,MKhammassi23,Vega-2023,Psomas-dec2023}. Recent results also looked into using multiple fluid antennas at one or both sides \cite{New-twc2023,XLai24} and employing FAS for security monitoring \cite{JYao24}. There has been great interest in FAS recently, producing many useful results. A relatively up-to-date list of recent literature on FAS can be found in \cite{LZhu24}.

On the other hand, reconfigurable intelligent surfaces (RIS) is widely considered as an enabling technology for the sixth generation (6G) \cite{WuQ1,KZhi221}. Unlike traditional relays, RIS does not require radio frequency (RF) links and amplifiers, which significantly reduces power consumption and hardware costs. Motivated by the above advantages, it is reasonable to combine RIS with FAS to boost the performance of IoT communication systems. However, the integration of FAS and RIS is not well understood and it remains an uncharted territory.

A major challenge in implementing FAS-RIS systems is the requirement for the base station (BS) to acquire instantaneous channel state information (CSI) to optimize the RIS reflection elements, which leads to considerable overhead \cite{JChu24,PSAung24}. To mitigate this, using statistical CSI instead to design the RIS phase shifters becomes an attractive alternative. This approach not only reduces the BS's burden and the feedback overhead \cite{MMZhao21,KZhi222} but also greatly lowers the computational complexity involved in adjusting the RIS reflection elements, since these adjustments are needed only at the intervals of statistical CSI updates rather than at each channel coherence time. %Thus, designing RIS reflection elements based on statistical CSI is not only practical but also enhances the performance of FAS-RIS systems.

This correspondence aims to investigate the FAS-RIS communication systems, by adopting the block-diagonal matrix approximation (BDMA) model in \cite{PR24}, analyzing the outage probability of the proposed system, and optimizing the phase shifting matrix of the RIS using the derived analytical results. The simulation results demonstrate the accuracy of the approximate outage probability based on the BDMA model, and that our proposed scheme outperforms significantly other benchmarks.

\vspace{-2mm}
\section{System Model}
Consider a FAS-RIS downlink system, consisting of a single fixed-position antenna BS, a RIS with $M$ reflection elements and a user that is equipped with a single $N$-port fluid antenna. The user dynamically switches the fluid antenna to the most desirable port over a linear space of size $W\lambda$, in which $W$ is the normalized length and $\lambda$ is the carrier wavelength \cite{KKWong21}. We assume that there is no direct link between the BS and user and the RIS is necessary to restore the communication link \footnote{Weak line-of-sight (LoS) links between the BS and the user may exist in certain scenarios, which is an interesting area for future research.}. %To facilitate communication between the BS and the user, the RIS is utilized to enable signal transmission from the BS to the user.

Given the fact that the RIS is usually installed above the ground and close to the BS, the BS-RIS channel is typically purely a line-of-sight (LoS) link \cite{KZhi222}. Accordingly, the BS-RIS channel can be modeled as
 \begin{align}\label{eq1}
\mathbf{g}= \sqrt{\beta}\bar{\mathbf{g}}\ \in{\mathbb{C}^{1\times M}},
\end{align}
where $\beta$ is the path-loss, and $\bar{\mathbf{g}}$ represents the LoS channel.

Then, since the distance between the RIS and the user may be large and the height of the user is limited, there exists several non-line-of-sight (NLoS) links between the RIS and the user. Consequently, the Rician channel model is adopted for the RIS-user channel. By denoting the channel from the RIS to the $k$-th port of the user as $\mathbf{h}_k\in{\mathbb{C}^{1\times{M}}}$, we have
 \begin{align}\label{eq2}
\mathbf{h}_k = \sqrt{\frac{\alpha K}{K+1}}\bar{\mathbf{h}}_k+\sqrt{\frac{\alpha}{K+1}}\tilde{\mathbf{h}}_k,
\end{align}
where $\alpha$ is the path loss, $K$ denotes the Rician factor, $\bar{\mathbf{h}}_k$ is the LoS component \footnote{The far-field model is considered in this paper, thus all ports in the FAS have the same LoS component, i.e., $\bar{\mathbf{h}}_1=\cdots=\bar{\mathbf{h}}_N=\bar{\mathbf{h}}$. Moreover, the LoS components are assumed to be constant, as th BS, the RIS, and the user are fixed. These components can be accurately obtained using channel estimation algorithms, such as the least squares algorithm \cite{PSAung24,MMZhao21}.},  $ \tilde{\mathbf{h}}_k$ is the NLoS component, whose elements are independently and identically distributed (i.i.d.) zero-mean complex Gaussian random variables (RVs) with unit variance.

Additionally, given the close proximity of the $N$ ports in the FAS, the channels $\mathbf{h}_k, \forall k$, are inherently correlated. We adopt the Jakes' model, where the correlation coefficient between the first port and the $k$-th port is modeled by \cite{KKWong20}
\begin{align}\label{eq4}
\mu_{1,k} &=J_0\left(\frac{2\pi(k-1)}{N-1}W\right), k\in\mathcal{N}=\{1,\dots,N\},
\end{align}
where $J_0(\cdot)$ is the zero-order Bessel function of the first kind. The correlation coefficient matrix of $[\mathbf{h}_1(m),\dots,\mathbf{h}_N(m)]$, in which the index $m$ specifies the $m$-th RIS element, is a Toeplitz matrix given by \cite{KKWong20,PR24}
\vspace{-10pt}\begin{align}\label{eq4a}
\mathbf{\Sigma}\in \mathbb{R}^{N\times N}  =\left(
         \begin{array}{cccc}
         \mu_{1,1} &\mu_{1,2} & \cdots & \mu_{1,N} \\
        \mu_{1,2} & \mu_{1,1} & \cdots & \mu_{1,N-1} \\
        \vdots &  & \ddots & \vdots \\
       \mu_{1,N}& \mu_{1,N-1} & \cdots &\mu_{1,1} \\
         \end{array}
       \right).
\end{align}

For both ease of analysis and accuracy, we use the BDMA model \cite{PR24} to approximate the spatial correlation structure. Specifically, we employ a block-diagonal correlation matrix to describe the spatial correlation structure of the FAS $N$ ports, which can be expressed as
\begin{equation}\label{eq3}
\hat{\bf\Sigma}\in \mathbb{R}^{N\times N}={\rm Blkdiag}(\mathbf{C}_1,\dots,\mathbf{C}_B),
\end{equation}
where ${\rm Blkdiag}(\cdot)$ returns a block diagonal matrix, with each submatrix $\mathbf{C}_b$ being a constant correlation matrix of size $L_b$ and having the correlation parameter $\mu^2_b$ as entries, i.e.,
\begin{equation}\label{eq3a}
\mathbf{C}_b\in \mathbb{R}^{L_b\times L_b}={\rm topelitz}(1, \mu_b^2,\dots, \mu_b^2),
\end{equation}
where $\sum\nolimits_{b=1}^{B}L_b=N$, and $L_b$ is found by \cite[Algorithm 1]{PR24}.

For $L_b$ ports in the $b$-th submatrix, the channels $\mathbf{h}_k$, for $k\in\{1,\dots,L_b\}$, are inherently correlated. The mathematical expression  can be given by
\begin{align}
\mathbf{h}_k =\sqrt{\frac{\alpha K}{K+1}}\bar{\mathbf{h}}_k+ \mu_b\tilde{\mathbf{h}}_b+ \sqrt{1-\mu_b^2}\mathbf{e}_k,
\end{align}
where  $\tilde{\mathbf{h}}_b$ and $\mathbf{e}_k$ are i.i.d.~RVs, following the distribution $\mathcal{CN}\left(\mathbf{0},\frac{\alpha}{K+1}\mathbf{I}_M\right)$ in which ${\bf I}_M$ is an identity matrix.

Further, let us denote $\mathbf{\Phi}=\mathrm{diag}\{\phi_1, \dots, \phi_m, \dots, \phi_{M}\}$ as the reflection matrix of the RIS, where $\phi_m=\exp(j\theta_m)$ for $m\in\mathcal{M}=\{1,\dots,M\}$ and $\theta_m\in(0,2\pi]$ is the phase shift of the $m$-th element. Accordingly, when
the BS sends a signal $x$ with $\mathbb{E}[|x|^2]=1$ to the user using the RIS, the received signal at the $k$-th port of the user can be expressed as
\begin{align}\label{eq5}
y_k &=\sqrt{P}\mathbf{h}_k\mathbf{\Phi}\mathbf{g}^Hx+n,
\end{align}
where $P$ represents the transmission power of the BS, $n\sim\mathcal{CN}(0,\sigma^2)$ denotes the additive Gaussian noise at the user.

\section{Outage Probability Analysis}
In this section, we provide a detailed analysis of the outage probability of the considered FAS-RIS systems. According to \eqref{eq5}, the signal-to-noise ratio (SNR) at the $k$-th port of the user can be expressed as
\begin{align}\label{eq6}
\gamma_{k}&=\frac{PA^2_k}{\sigma^2},
\end{align}
where $A_k=|\mathbf{h}_k\mathbf{\Phi}\mathbf{g}^H|$. Next, let us further define the maximum of $A_k$  as $A_{\max} =\max\{A_1, A_2, \dots, A_N\}$.  As a result, the SNR of the FAS-aided user is given by
\begin{align}\label{eq8}
\gamma&=\frac{P|A_{\max}|^2}{\sigma^2}.
\end{align}
Building upon this, when giving the transmission rate as $R$, the outage probability at the user can be defined as
\begin{align}\label{eq9}
\mathbb{P}^{\mathrm{out}}&=\mathbb{P}(\log_2(1+\gamma)<R).
\end{align}

\subsection{Outage Probability Based on BDMA Model}
When the RIS only has the knowledge of statistical CSI, it can be assumed that the phase $\phi_m$ remains constant in several transmission durations. For the $b$-th submatrix of $\hat{\bf\Sigma}$, $\mathbf{h}_k\mathbf{\Phi}\mathbf{g}^H$ can be rewritten as
\begin{align}
\mathbf{h}_k\mathbf{\Phi}\mathbf{g}^H=&\sqrt{\frac{\alpha K}{K+1}} \bar{\mathbf{h}}_k\mathbf{\Phi}\mathbf{g}^H+\mu \tilde{\mathbf{h}}_b\mathbf{\Phi}\mathbf{g}^H  + \sqrt{1-\mu^2}\mathbf{e}_k\mathbf{\Phi}\mathbf{g}^H\nonumber\\
=&\eta+\mu\sum_{i=1}^M\tilde{h}_b^ie^{j\theta_m}g^i+\sqrt{1-\mu^2}\sum_{i=1}^Me_k^ie^{j\theta_m}g^i,\label{Bq1}
\end{align}
where $\eta=\sqrt{\frac{\alpha K}{K+1}} \bar{\mathbf{h}}_k\mathbf{\Phi}\mathbf{g}^H$, and the scalars in (\ref{Bq1}) denote the corresponding entries of the matrix and vector variables.

We can find that $\tilde{h}_b^ie^{j\theta_m}g^i, i\in\mathcal{M}$ is an i.i.d.~RV. Then, by using central limit theorem (CLT), we have \cite{BKnaeble}
\begin{align}\label{Bq2}
\tilde{\mathbf{h}}_b\mathbf{\Phi}\mathbf{g}^H \sim \mathcal{CN}(0,\bar{\sigma}^2),
\end{align}
where
\begin{align}\label{Bq3}
\bar{\sigma}^2=\sum_{i=1}^M\mathbb{V}\left[h_0^ie^{j\theta_i}g^i\right]=\frac{M\alpha\beta}{K+1},
\end{align}
where $\mathbb{V}$ denotes the variance computation.

Similarly, $\mathbf{e}_k\mathbf{\Phi}\mathbf{g}^H$ is the sum of $e_k^ie^{j\theta_m}g^i, i\in\mathcal{M}$, which is an i.i.d.~RV. Thus, we also have the distribution as
\begin{align}\label{Bq4}
\mathbf{e}_k\mathbf{\Phi}\mathbf{g}^H\sim \mathcal{CN}(0,\bar{\sigma}^2).
\end{align}

As a consequence, given $\tilde{\mathbf{h}}_b\mathbf{\Phi}\mathbf{g}^H$ and with the definition $\Lambda_b \triangleq \left|\eta+\mu\tilde{\mathbf{h}}_b\mathbf{\Phi}\mathbf{g}^H\right|$, the probability density function (PDF) of $A_k$ can be expressed as
\begin{align}\label{Bq8}
f_{A_k | \Lambda_b}(r_k | r_b) = &\frac{2r_k}{\bar{\sigma}^2 (1-\mu^2)} e^{-\frac{r_k^2+r_b^2}{\bar{\sigma}^2 (1-\mu^2)}} I_{0}\left(\frac{2r_kr_b}{\bar{\sigma}^2 (1-\mu^2)}\right),
\end{align}
where $I_{0}(u)$ denotes the modified Bessel function of the first kind and order zero. Its series representation is given by
\begin{align}\label{Bq9}
I_0(z) = \sum_{k=0}^{\infty} \frac{z^{2k}}{2^{2k} k!\Gamma(k+1)},
\end{align}
where $\Gamma(k+1)=k!$.

Conditioned on $\Lambda_b$, it is clear that $A_1, \dots, A_N$ are mutually independent. This then allows us to deduce the joint PDF of $A_1, \dots, A_N$ as
\vspace{-10pt}\begin{align}\label{Bq10}
&f_{A_1, \dots, A_N | \Lambda_b}(r_1, \dots, r_N| r_b)\nonumber\\
&=\prod_{k\in \mathcal{K}_b}\frac{2r_k}{\bar{\sigma}^2 (1-\mu^2)} e^{-\frac{r_k^2+r_b^2}{\bar{\sigma}^2 (1-\mu^2)}} I_{0}\left(\frac{2r_kr_b}{\bar{\sigma}^2 (1-\mu^2)}\right).
\end{align}
Besides, given the fact that $\Lambda_b$ follows the Rician distribution, the PDF of $\Lambda_b$ is given by
\begin{align}\label{Bq10A}
f_{\Lambda_b}(r_b)=&\frac{2r_b}{\bar{\sigma}^2 \mu^2} e^{-\frac{r_b^2+|\eta|^2}{\bar{\sigma}^2 \mu^2}} I_{0}\left(\frac{2r_b|\eta|}{\bar{\sigma}^2 \mu^2}\right).
\end{align}
Consequently, by combining \eqref{Bq10} with \eqref{Bq10A}, the joint PDF of $\Lambda_b, A_1, \dots, A_N$ is written as
\vspace{-10pt}\begin{align}\label{Bq11}
&f_{\Lambda_b, A_1, \dots, A_N}(r_b, \dots, r_N)\nonumber\\
&=\frac{2r_b}{\bar{\sigma}^2 \mu^2} e^{-\frac{r_b^2+|\eta|^2}{\bar{\sigma}^2 \mu^2}} I_{0}\left(\frac{2r_b|\eta|}{\bar{\sigma}^2 \mu^2}\right) \nonumber\\ & \ \times \prod_{k\in \mathcal{K}_b}\frac{2r_k}{\bar{\sigma}^2 (1-\mu^2)} e^{-\frac{r_k^2+r_b^2}{\bar{\sigma}^2 (1-\mu^2)}}
\times I_{0}\left(\frac{2r_kr_b}{\bar{\sigma}^2 (1-\mu^2)}\right).
\end{align}
Building upon this and   \cite[(32)]{KKWong20}, the joint cumulative distribution function (CDF) of $\Lambda_b, A_1, \dots, A_N$ is formulated as
\hspace{-15pt}\begin{align}\label{Bq12}
F&_{\Lambda_b, A_1, A_2, \cdots, A_N}(r_b, r_1, \dots, r_N)\nonumber\\
=&\int_0^\infty\frac{2r_b}{\bar{\sigma}^2 \mu^2} e^{-\frac{r_b^2+|\eta|^2}{\bar{\sigma}^2 \mu^2}} I_{0}\left(\frac{2r_b|\eta|}{\bar{\sigma}^2 \mu^2}\right)\times \nonumber\\
& \prod_{k\in \mathcal{K}_b}\left[1-Q_1\left(\sqrt{\frac{2}{\bar{\sigma}^2(1-\mu^2)}}r_b,\sqrt{\frac{2}{\bar{\sigma}^2(1-\mu^2)}}r_k\right)\right]dr_b.
\end{align}
Therefore, we can obtain the joint CDF of $\{A_n\}$ as
\begin{align}\label{Bq13}
F&_{\{A_n\}}(r_1, \dots, r_N)\nonumber\\
=&\prod_{b=1}^B\int_0^\infty\frac{2r_b}{\bar{\sigma}^2 \mu^2} e^{-\frac{r_b^2+|\eta|^2}{\bar{\sigma}^2 \mu^2}} I_{0}\left(\frac{2r_b|\eta|}{\bar{\sigma}^2 \mu^2}\right)\times \nonumber\\
& \prod_{k\in \mathcal{K}_b}\left[1-Q_1\left(\sqrt{\frac{2}{\bar{\sigma}^2(1-\mu^2)}}r_b,\sqrt{\frac{2}{\bar{\sigma}^2(1-\mu^2)}}r_k\right)\right]dr_b,
\end{align}
where $Q_1(\cdot,\cdot)$ is the first-order Marcum $Q$-function.

As such, the outage probability can be found by substituting $r_1=\cdots=r_N=\sqrt{\gamma_{\rm th}}$ into the joint CDF \eqref{Bq13}, given by
\hspace{-15pt}\begin{align}\label{Bq14}
\mathbb{P}^{\mathrm{out}}=&\prod_{b=1}^B\int_0^\infty\frac{2r_b}{\bar{\sigma}^2 \mu^2} e^{-\frac{r_b^2+|\eta|^2}{\bar{\sigma}^2 \mu^2}} I_{0}\left(\frac{2r_b|\eta|}{\bar{\sigma}^2 \mu^2}\right)\times\nonumber\\
& \left[1-Q_1\left(\sqrt{\frac{2}{\bar{\sigma}^2(1-\mu^2)}}r_b,\sqrt{\frac{2\gamma_{\rm th}}{\bar{\sigma}^2(1-\mu^2)}}
\right)\right]^{L_b}dr_b,
\end{align}
where $\gamma_{\rm th}=(2^R-1)\sigma^2/P$.

\subsection{Upper Bound}
According to the findings in \cite{PR24}, the outage probability derived using the independent antennas equivalent (IAE) model\footnote{The IAE model simplifies the analysis by setting $\mu_b$ to 1, i.e., $\mu_b\rightarrow 1$.} serves as an upper bound to that obtained using the BDMA model when $L_b$, for $b \in {1, \dots, B}$, remains fixed. Specifically, when $\mu_b=1$, we find that
\begin{align}\label{Dq1}
\mathbf{h}_b\mathbf{\Phi}\mathbf{g}^H=&\sqrt{\frac{\alpha K}{K+1}}\bar{\mathbf{h}}_b\mathbf{\Phi}\mathbf{g}^H+\tilde{\mathbf{h}}_b\mathbf{\Phi}\mathbf{g}^H.
\end{align}
We employ the CLT to establish that the product of the channel vectors $\mathbf{h}_b$, the matrix $\mathbf{\Phi}$, and the vector $\mathbf{g}$ follows a complex Gaussian distribution, i.e.,
\begin{align}\label{Dq2}
\mathbf{h}_b\mathbf{\Phi}\mathbf{g}^H\sim \mathcal{CN}\left(\eta,\bar{\sigma}^2\right).
\end{align}

Consequently,  $|A_b|, b\in\mathcal{B}$ follows the Rician distribution, whose PDF is given by
\begin{align}\label{Dq3}
f_{|A_b|}(r_b) = \frac{2r_b}{\bar{\sigma}^2} e^{-\frac{r_b^2+|\eta|^2}{\bar{\sigma}^2}} I_{0}\left(\frac{2r_b|\eta|}{\bar{\sigma}^2}\right).
\end{align}
Thus, we can derive the joint CDF of $|A_1|, |A_2|, \dots, |A_B|$ as
\begin{align}\label{Dq4}
F_{|A_1|, |A_2|, \dots, |A_B|}&(r_1, \dots, r_B)\nonumber\\
&=\prod_{b=1}^B \left[1-Q_1\left(\sqrt{\frac{2}{\bar{\sigma}^2}}|\eta|,\sqrt{\frac{2r_b}{\bar{\sigma}^2}}\right)\right].
\end{align}
Hence, the outage probability can be found by substituting $r_1=\cdots=r_B=\gamma_{\rm th}$ into \eqref{Dq4}, which is given by
\begin{align}\label{Dq5}
\check{\mathbb{P}}^{\mathrm{out}}=\left[1-Q_1\left(\sqrt{\frac{2}{\bar{\sigma}^2}}|\eta|,\sqrt{\frac{2\gamma_{\rm th}}{\bar{\sigma}^2}}\right)\right]^B.
\end{align}

\subsection{Lower Bound and Asymptotic Analysis}
From \eqref{Bq8}, we have the lower bound of the PDF of $A_k$ as
\begin{align}\label{Bq15}
\hat{f}_{A_k | \Lambda_b}(r_k | r_b) = &\frac{2r_k}{\bar{\sigma}^2 (1-\mu^2)} e^{-\frac{r_k^2+r_b^2}{\bar{\sigma}^2 (1-\mu^2)}}.
\end{align}
Considering $\Lambda_b$, we note that the variables $A_1, \dots, A_N$ are mutually independent. This independence allows us to deduce the lower bound of the joint PDF of $A_1, \dots, A_N$ as
\begin{align}\label{Bq16}
\hat{f}_{A_1, \dots, A_N | \Lambda_b}(r_1, \dots, r_N| r_b)=\prod_{k\in \mathcal{K}_b}\frac{2r_k}{\bar{\sigma}^2 (1-\mu^2)} e^{-\frac{r_k^2+r_b^2}{\bar{\sigma}^2 (1-\mu^2)}}.
\end{align}

Similarly, the low bound of the PDF of $\Lambda_b$ is expressed as
\begin{align}\label{Bq17}
\hat{f}_{\Lambda_b}(r_b)=&\frac{2r_b}{\bar{\sigma}^2 \mu^2} e^{-\frac{r_b^2+|\eta|^2}{\bar{\sigma}^2 \mu^2}}.
\end{align}
Utilizing \eqref{Bq16} and \eqref{Bq17},  we can derive the lower bound of the joint PDF for $\Lambda_b, A_1, \dots, A_N$ as
\vspace{-5pt}\begin{align}\label{Bq18}
&\hat{f}_{\Lambda_b, A_1, \dots, A_N}(r_b, \dots, r_N)\nonumber\\
&=\frac{2r_b}{\bar{\sigma}^2 \mu^2} e^{-\frac{r_b^2+|\eta|^2}{\bar{\sigma}^2 \mu^2}}
\times \prod_{k\in \mathcal{K}_b}\frac{2r_k}{\bar{\sigma}^2 (1-\mu^2)} e^{-\frac{r_k^2+r_b^2}{\bar{\sigma}^2 (1-\mu^2)}}.
\end{align}
Following this, by employing \eqref{Bq18}, the lower bound of the outage probability $\mathbb{P}^{\mathrm{out}}$ is obtained by \eqref{Bq19},
\begin{figure*}[ht]
\centering
\begin{align}\label{Bq19}
\hat{\mathbb{P}}^{\mathrm{out}}_{m}
&\hspace{.5mm}=e^{-\frac{B|\eta|^2}{\bar{\sigma}^2 \mu^2}} \prod_{b=1}^B\int_{0}^{\infty} \frac{2r_b}{\bar{\sigma}^2 \mu^2} e^{-\frac{r_b^2}{\bar{\sigma}^2 \mu^2}} \prod_{k\in \mathcal{K}_b}e^{-\frac{r_b^2}{\bar{\sigma}^2 (1-\mu^2)}} \left[1-e^{-\frac{\gamma_k^2}{\bar{\sigma}^2 (1-\mu^2)}}\right]dr_b\notag\\
&\overset{(a)}{=}e^{-\frac{B|\eta|^2}{\bar{\sigma}^2 \mu^2}} \prod_{b=1}^B \left[1-e^{-\frac{\gamma_{th}}{\bar{\sigma}^2 (1-\mu^2)}}\right]^{L_b} \int_{0}^{\infty} \frac{2r_b}{\bar{\sigma}^2 \mu^2} e^{-\frac{r_b^2}{\bar{\sigma}^2 \mu^2}} e^{-\frac{L_br_b^2}{\bar{\sigma}^2 (1-\mu^2)}} dr_b =e^{-\frac{B|\eta|^2}{\bar{\sigma}^2 \mu^2}} \prod_{b=1}^B \frac{1-\mu^2}{1+(L_b-1)\mu^2}\left[1-e^{-\frac{\gamma_{th}}{\bar{\sigma}^2 (1-\mu^2)}}\right]^{L_b}
\end{align}
\hrulefill
\end{figure*}
appearing at the top of next page. Here, step $(a)$ uses $r_1=\cdots=r_N=\sqrt{\gamma_{\rm th}}$.

From \eqref{Bq19}, by using the approximation $e^{-x}=1-x$ for tiny value $|x|$, we derive the asymptotic expression for the outage probability in the high SNR regime as
\vspace{-10pt}\begin{align}\label{Bq20}
\mathbb{P}^{\mathrm{out}}_{m}\simeq \prod_{b=1}^B \eta_b \left[\frac{\gamma_{\rm th}}{\bar{\sigma}^2(1-\mu^2)}\right]^{L_b},
\end{align}
where $\eta_b = \frac{(1-\mu^2)e^{-\frac{|\eta|^2}{\bar{\sigma}^2 \mu^2}}}{1+(L_b-1)\mu^2}$.

\emph{Remark 1}: The asymptotic outage probability in \eqref{Bq20} indicates that the diversity order of the system is $\sum\nolimits_{b=1}^{B}L_b=N$.
\vspace{-10pt}
\vspace{-10pt}\section{Outage Probability Optimization}
In this section, our objective is to minimize the outage probability at the user under the reflection coefficient constraint of the RIS, as formulated below:
\vspace{-5pt}\begin{align}\label{cq6}
\min_{\mathbf{\Phi}} \ \mathbb{P}^{\mathrm{out}}~~\mbox{s.t.}~~|\phi_m|=1,\ \forall m\in\mathcal{M}.
\end{align}
Given the complexity of the expression for $\mathbb{P}^{\mathrm{out}}$, we begin by approximating the outage probability of the system. To do so, we simplify the expression by setting $I_0(z)$ to 1. Based on \eqref{Bq14}, the approximation of $\mathbb{P}^{\mathrm{out}}$ is expressed as
\vspace{-5pt}\begin{align}\label{dq1}
\hat{\mathbb{P}}^{\mathrm{out}}=&e^{-\frac{B\eta^2}{\bar{\sigma}^2 \mu^2}} \prod_{b=1}^B\int_0^\infty\frac{2r_b}{\bar{\sigma}^2 \mu^2} e^{-\frac{r_b^2}{\bar{\sigma}^2 \mu^2}} \times\nonumber\\
& \left[1-Q_1\left(\sqrt{\frac{2}{\bar{\sigma}^2(1-\mu^2)}}r_b,\sqrt{\frac{2\gamma_{\rm th}}{\bar{\sigma}^2(1-\mu^2)}}\right)\right]^{L_b}dr_b.
\end{align}
Consequently, \eqref{cq6} can be reformulated as
\vspace{-5pt}\begin{align}\label{dq2}
\min_{\mathbf{\Phi}}\ \ \hat{\mathbb{P}}^{\mathrm{out}}\ \ \mbox{s.t.}\  |\phi_m|=1,\ \forall\ m\in\mathcal{M}.
\end{align}

From \eqref{dq1}, we observe that $\hat{\mathbb{P}}^{\mathrm{out}}$ decreases as $\eta$ increases. Given that $\mu$ is constant, Problem \eqref{dq2} can be equivalently reformulated as
\vspace{-10pt}\begin{align}\label{dq3}
\max_{\mathbf{\Phi}} \  \left|\bar{\mathbf{h}}\mathbf{\Phi}\mathbf{g}^H\right|~~\mbox{s.t.}~~|\phi_m|=1,\ \forall m\in\mathcal{M}.
\end{align}
According to \cite{WuQ1}, the optimal $\phi_m$ is given by
\vspace{-2pt}\begin{align}\label{dq4}
\phi_m^o=e^{-j(\angle\bar{\mathbf{h}}(m)-\angle\mathbf{g}(m))}.
\end{align}

\vspace{-5pt}
\emph{Remark 2}: The solution in \eqref{dq4} indicates that the reflection elements of the RIS are strategically designed to maximize the LoS components of the cascaded BS-RIS-user channel.

\section{Numerical Results}
In our simulations, we model the scenario within a three-dimensional (3D) coordinate system. The BS, the RIS, and the user are located at $(0, 0, 5)$ m, $(15, 15, 5)$ m, and $(50, 0, 0)$ m, respectively. The channel from the BS to the RIS are characterized as distance-dependent flat Rician fading channel, with a large-scale path loss exponent of $2.2$ and a Rician factor of $1$ \cite{KZhi222}. The angles in the LoS channels are randomly generated within the range of $[0, 2\pi]$. The noise power is set at $\sigma^2 = -104$ dBm. We also assume the FAS size as $W = 5$, and a transmission rate from the BS of $5$ bps/Hz. Moreover, the correlation coefficient for each block is uniformly set at $\mu_{1}^2=\cdots =\mu_{b}^2=\cdots=\mu_{B}^2=0.97$ \cite{PR24}.  Monte Carlo simulations are conducted using the correlation coefficient matrix in \eqref{eq3}, with an averaging over $100,000$ iterations.

%\begin{figure*}[t]
%\centering
%\begin{minipage}{0.32\linewidth}
%{\includegraphics[width=1.0 \textwidth]{fig2.eps}}
%\caption{Outage probability versus $P$, where $N=50$. \vspace{-1ex}}\label{fig:oVSp}
%\end{minipage}
%\begin{minipage}{0.32\linewidth}
%{\includegraphics[width=1.0 \textwidth]{fig3.eps}}
%\caption{Outage probability versus $N$. }\label{fig:oVSn}
%\end{minipage}
%\begin{minipage}{0.32\linewidth}
% {\includegraphics[width=1.01\textwidth]{fig4.eps}}
%\caption{Outage probability based on BDMA model versus $P$; comparison of two reflection elements schemes, where $N=50$ and $M=9$. \vspace{-3ex}}\label{fig:oVSp2}
%\end{minipage}
%\end{figure*}

\begin{figure}[t]
\begin{centering}
\centering{}\includegraphics[scale=0.39]{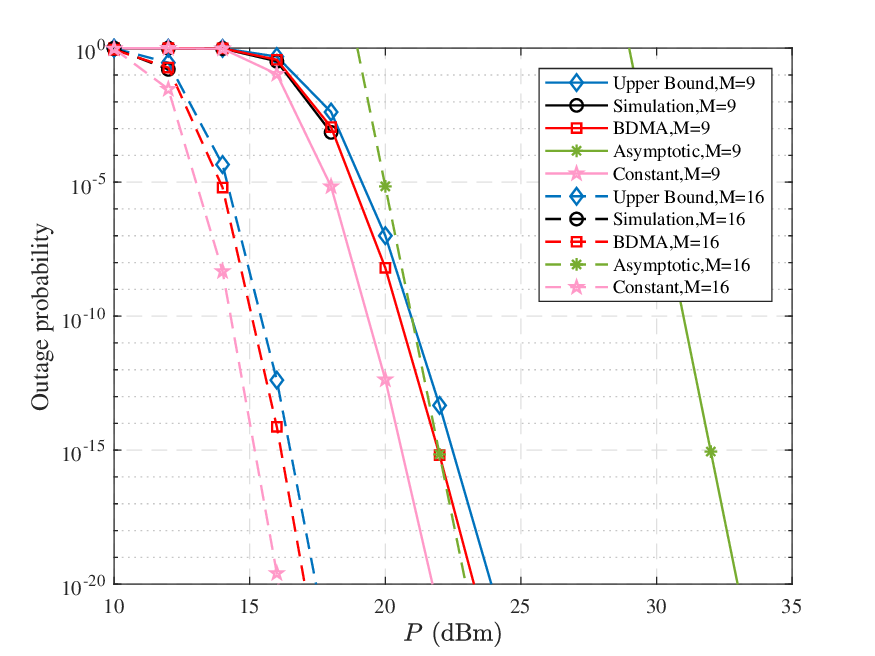}
\caption{Outage probability versus $P$, where $N=50$. \vspace{-1ex}}\label{fig:oVSp}
\par\end{centering}
\begin{centering}
\noindent \includegraphics[scale=0.39]{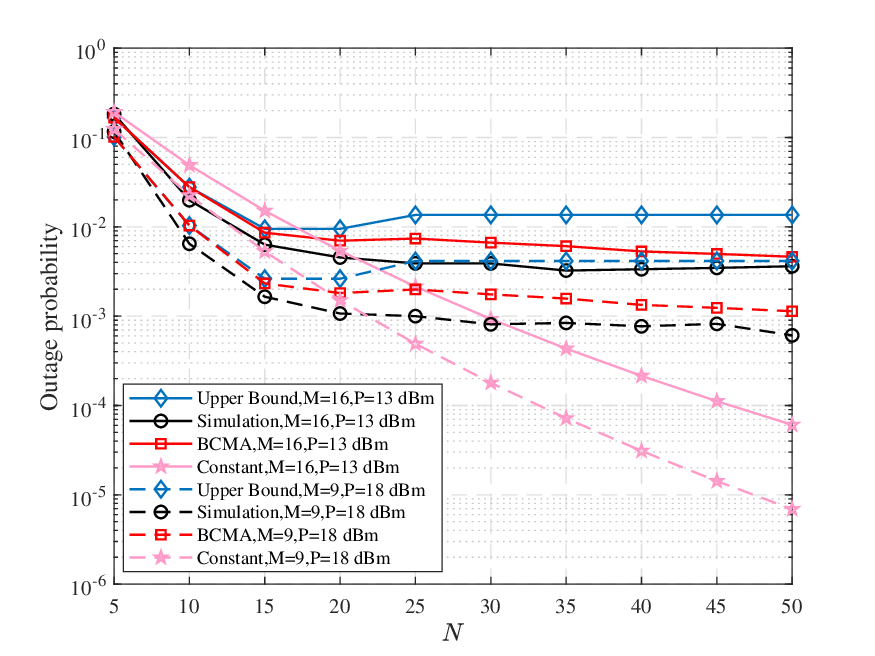}
\caption{Outage probability versus $N$. }\label{fig:oVSn}
\par\end{centering}
\begin{centering}
\noindent \includegraphics[scale=0.39]{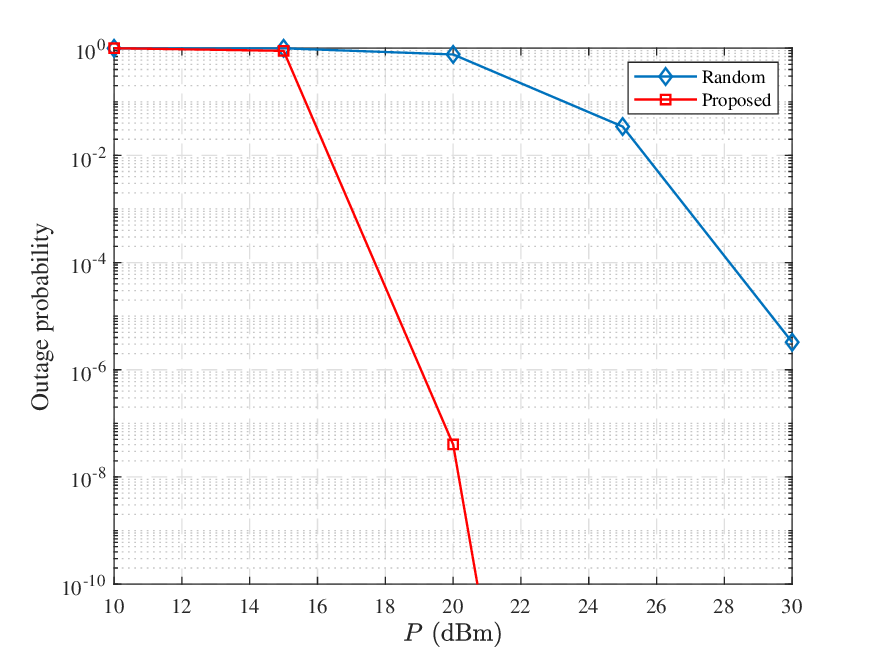}
\par\end{centering}
\caption{Outage probability based on BDMA model versus $P$; comparison of two reflection elements schemes, where $N=50$ and $M=9$.}\label{fig:oVSp2}
\vspace{-3mm}
\end{figure}

In Fig.~\ref{fig:oVSp}, we investigate the impact of the transmit power $P$ on the outage probability when $N=50$. In the legend, ``BDMA'' denotes the theoretical analysis based on the BDMA model, while ``Simulation" corresponds to the Monte Carlo simulation results; ``Upper Bound'' represents the theoretical analysis using the IAE model; ``Asymptotic'' is the asymptotic analysis, and ``Constant'' denotes the approximation model based on \cite{KKWong20}. As shown in Fig.~\ref{fig:oVSp}, the outage probabilities for all schemes decrease as $P$  increases, highlighting the typical inverse relationship between power and outage probability. Further analysis reveals that the ``BDMA" lines closely align with the simulated results, indicating that the BDMA model effectively captures the characteristics of the true correlation matrix as in \eqref{eq4a}. Additionally, it is observed that the outage probability decreases with an increase in the number of reflection elements $M$. Specifically, at $P=14$ dBm, the outage probability of ``BDMA'' with $M=9$ is about $0.955$, whereas it drastically reduces to about $7.41\times10^{-15}$  with  $M=16$. This significant decrease is attributed to the larger $M$ providing a higher DoF, to manage and mitigate signal fading.

In Fig.~\ref{fig:oVSn}, we demonstrate the impact of $N$ on the outage probability. As seen from Fig.~\ref{fig:oVSn}, the outage probabilities of all schemes decrease with the increasing of $N$. This is because the larger $N$, the larger the SNR of the user. From Fig.~\ref{fig:oVSn}, we can also observe that the larger $N$, the more accurate  the outage performance of ``BCMA". This improvement is likely because the block-correlation matrix approximation benefits from asymptotic statistical results which become more reliable as $N$ increases. Moreover, the discrepancy between the ``Simulation'' and ``Constant'' results widens as $N$ increases, highlighting that the constant model becomes less effective for larger $N$.  This indicates that the assumptions underlying the constant model may not hold well under the conditions of high $N$, leading to deviations from the Monte Carlo results.

In Fig.~\ref{fig:oVSp2}, we study the impact of the transmit power $P$ on the outage probabilities of two reflection elements schemes, where $N=50$ and $M=9$. In the legend, ``Random'' denotes the phase of each reflection element of the RIS is randomly selected that follows a uniform distribution between 0 and $2\pi$, while ``Proposed'' corresponds to the solution proposed in \eqref{dq4}. From Fig.~\ref{fig:oVSp2}, we can find that the proposed scheme always outperforms the ''Random'' scheme.

\section{Conclusion}
This correspondence investigated the integrated FAS-RIS model where there is no direct link between the BS and user. The BDMA model was adopted to permit our outage probability analysis. Our contributions include the upper bound, lower bound, and asymptotic approximation of the outage probability. Besides, we used the analytical outage probability expression for designing the passive beamforming of the RIS. Our numerical results clearly demonstrated the accuracy of the approximation in the analysis, and that the outage probability performance of our proposed scheme surpasses existing benchmarks. Moreover, it remains an interesting future work to explore the scenario where exists the LoS links between BS and user.

\end{document}